\documentclass[conference, a4paper]{IEEEtran}
\IEEEoverridecommandlockouts
\usepackage{cite}
\usepackage{amsmath,amssymb,amsfonts}
\usepackage{algorithmic}
\usepackage{graphicx}
\usepackage{textcomp}
\usepackage{xcolor}
\usepackage[left=1.62cm,right=1.62cm,top=1.9cm,bottom=4.4cm]{geometry}
\usepackage[singlespacing]{setspace} 
\def\BibTeX{{\rm B\kern-.05em{\sc i\kern-.025em b}\kern-.08em
    T\kern-.1667em\lower.7ex\hbox{E}\kern-.125emX}}

\usepackage[mathlines,switch]{lineno}

\makeatletter

\let\old@ps@headings\ps@headings
\let\old@ps@IEEEtitlepagestyle\ps@IEEEtitlepagestyle
\def\confheader#1{%
    \def\ps@headings{%
        \old@ps@headings%
        \def\@oddhead{\strut\hfill#1\hfill\strut}%
        \def\@evenhead{\strut\hfill#1\hfill\strut}%
    }%
    \def\ps@IEEEtitlepagestyle{%
        \old@ps@IEEEtitlepagestyle%
        \def\@oddhead{\strut\hfill#1\hfill\strut}%
        \def\@evenhead{\strut\hfill#1\hfill\strut}%
    }%
    \ps@headings%
}
\makeatother

\title{Deep learning classification system for coconut maturity levels based on acoustic signals
\thanks{This work was funded by the ERDT program under the DOST, Republic of the Philippines.}

}

\author{\IEEEauthorblockN{June Anne Caladcad}
    \IEEEauthorblockA{\textit{Department of Industrial Engineering} \\
    \textit{University of San Jose-Recoletos}\\
    Cebu City 6000 Philippines \\
    juneanne.caladcad@usjr.edu.ph}
    \and
    \IEEEauthorblockN{Eduardo Jr Piedad}
    \IEEEauthorblockA{\textit{Advanced Science and Technology Institute} \\
    \textit{Department of Science and Technology}\\
    Quezon City 1101 Philippines \\
    eduardojr.piedad@asti.dost.gov.ph}
}

\IEEEoverridecommandlockouts
\usepackage{tikz}
\usepackage{textcomp}
\usepackage{hyperref}
\usepackage{lipsum}

\newcommand\copyrighttext{%
  \footnotesize \textcopyright 2024 IEEE. Personal use of this material is permitted. Permission from IEEE must be obtained for all other uses, in any current or future media, including reprinting/republishing this material for advertising or promotional purposes, creating new collective works, for resale or redistribution to servers or lists, or reuse of any copyrighted component of this work in other works.  
}
\newcommand\copyrightnotice{%
\begin{tikzpicture}[remember picture,overlay]
\node[anchor=south,yshift=10pt] at (current page.south) {\fbox{\parbox{\dimexpr\textwidth-\fboxsep-\fboxrule\relax}{\copyrighttext}}};
\end{tikzpicture}%
}

\begin{document}

\maketitle
\copyrightnotice

\begin{abstract}
The advancement of computer image processing, pattern recognition, signal processing, and other technologies has gradually replaced the manual methods of classifying fruit with computer and mechanical methods. In the field of agriculture, the intelligent classification of post-harvested fruit has enabled the use of smart devices that creates a direct impact on farmers, especially on export products. For coconut classification, it remains to be traditional in process. This study presents a classification of the coconut dataset based on acoustic signals. To address the imbalanced dataset, a data augmentation technique was conducted through audiomentation and procedural audio generation methods. Audio signals under premature, mature, and overmature now have 4,050, 4,050, and 5,850 audio signals, respectively. To address the updation of the classification system and the classification accuracy performance, deep learning models were utilized for classifying the generated audio signals from data generation. Specifically, RNN and LSTM models were trained and tested, and their performances were compared with each other and the machine learning methods used by Caladcad et al. (2020). The two DL models showed impressive performance with both having an accuracy of 97.42\% and neither of them outperformed the other since there are no significant differences in their classification performance. 
\end{abstract}

\begin{IEEEkeywords}
Acoustic signal, RNN, LSTM, deep learning, data augmentation
\end{IEEEkeywords}

\section{Introduction}
Coconuts are cultivated in different regions around the world, specifically in tropical and sub-tropical countries \cite{1,2}. Among these countries, Indonesia has the world’s largest production, followed by the Philippines and then India \cite{1}. In 2018, the three largest coconut-producing countries—Indonesia, the Philippines, and India—contributed 78\% of the global coconut production and the remaining 12\% were from the rest of the coconut-producing countries \cite{1}. In the Philippines, coconut is the largest employer of agricultural land and labor \cite{3}. Coconuts occupy 23\% of the country’s total land devoted to agricultural use. There are a total of 2.5 million coconut farmers, consisting of smallholders who own an average farm size of 0.5–5 hectares while the rest are inheritors of industrial plantations \cite{3,4}. Furthermore, the Philippines has an annual coconut production of 14.7 metric tons in nut terms and has average export earnings of 91.4 billion PHP (1.7 billion USD) from 2014 to 2018 \cite{4}. About 25\% to 33\% of the country’s population is dependent on the coconut industry for their livelihood \cite{2}.

Different commercial purposes of coconuts vary on the classification of the fruit’s maturity level \cite{5}. Terdwongworakul et al. \cite{5} stated that fruit quality is strongly affected by the maturity level of coconut fruit. There are different levels of coconuts identified but most commonly, they are classified into three levels, i.e., premature, mature, and overmature \cite{5,6,7}. The classification of fruit is essentially the fundamental method of determining their maximum economic value \cite{8}. However, the classification of fruit has significant challenges and is proven to be a complex problem due to interclass similarities and irregular intraclass characteristics \cite{9}.

Traditionally, coconuts are sorted into their maturity levels manually. Farmers or traders would either use their fingernails, knuckles, or the blunt end of a knife to tap the coconuts and assess the sounds produced by them \cite{6}. However, the traditional process of manual classification poses a lot of drawbacks from human-related constraints such as inconsistency, time-consuming, variability, and subjectivity \cite{10}. Dealing with a large volume of coconuts to be exported or delivered for industrial processing as a fresh product, the traditional technique of manually sorting fruit will no longer be feasible \cite{6}. Behera et al. \cite{11} reported that due to the lack of skilled workers and human subjectivity in classifying coconuts, 30\% to 35\% of the harvested coconuts are wasted.

With the advancement of computer image processing, pattern recognition, and other technologies, manual fruit classification is gradually being replaced with mechanical methods \cite{8}. Specifically, the progression in artificial intelligence (AI) applications enabled the use of smart devices in the agricultural industry \cite{12}. Several methods were used in the literature to advance and improve the current classification system of coconuts. These include Terdwongworakul et al. \cite{5} using a simple linear regression model based on the resonant frequency, Hahn \cite{13} designing an on-line detector based on dissolved oxygen, pH values, and water volume using coconut water with a destructive technique, and Parvathi and Selvi \cite{14} using faster R-CNN algorithm to detect coconuts from real-time images and Google-searched images. While these offer significant insights and have promising results, the models may be perceived to be insufficient and impractical to be used for mass exportation.

While a significant effort has been made to study classification opportunities in sorting coconuts according to their maturity levels, limited effort has been made in designing a non-destructive classification system for coconut fruit. Creating an automated classification system can benefit the agricultural industry and coconut exportation companies in terms of saving time and money, improving accuracy and performance in sorting, and significantly lesser fruit would go to waste brought from human subjective perception. Thus, this paper attempted to fill in such gaps. First, this study applied data generation techniques to increase the data samples in the dataset of Caladcad et al. \cite{15}. Data augmentation has become a standard technique used in several studies to improve the performance of neural networks \cite{16}. Second, the performance accuracy was also attempted to be improved with the use of two major time-series DL algorithms, which are Recurrent Neural Network (RNN), and Long Short-Term Memory (LSTM). These algorithms were assessed based on their accuracy performance to see which is the best-fit model for the classification system. RNN has been successfully used in time series data and mining patterns in long-time sequences \cite{17,18}. While LSTM is most notable for sequence learning \cite{19,20}.

\section{Materials and Methods}
\subsection{Coconut dataset}

This study will be using the same coconut dataset used by the study of Caladcad et al. \cite{15}, which is available in the supplementary material provided by Caladcad et al. \cite{21}. The primary subjects are whole coconuts classified under the tall coconut variety in their post-harvest period. They are classified primarily into three maturity levels: “premature,” “mature,” and “overmature.” There is a total of 129 coconuts randomly gathered with 8, 36, and 85 coconuts for premature, mature, and overmature levels, respectively. Acoustic signals are gathered using a mechanized coconut tapper that synchronizes the tapping and recording process. This is to ensure uniformity across the dataset. The dataset, however, was limited by the imbalance of the distribution of samples per maturity levels due to the seasonal cycle of coconuts, thus, there was a clear domination of the number of overmature samples in the dataset. This also limits the study from having sufficient data for testing phase before data augmentation.

Data augmentation was applied to increase the amount of dataset. The implementation of data augmentation is summarized with the following steps: (1) pre-processing, (2) data generation, and (3) data validation. The pre-processing involves initial data cleaning, in which missing and wrongly labelled data were removed. Signals are then combined and feature to be extracted is selected. The combining and feature extraction methods explored are summarized in Table \ref{tab1}. These methods are assessed as to which methods can distinctly classify the data points to three classifications. It showed that using the summing method and Mel Frequency Cepstral Coefficient (MFCC) has the best results. Acoustic signals need to be combined since in the study of Caladcad et al. \cite{15}, three signals were extracted per coconut sample and data clustering is not possible for data per ridge knock. Finally, data were cleaned of any outliers.

For data augmentation, both audiomentation and procedural generation methods were used. Audiomentation manipulates features of the audio signal to generate more data \cite{22}, while procedural generation uses frequency features to create more data \cite{23}. Summarized in Table \ref{tab2} are the deformation techniques and filters used.

The generated dataset was then validated using a simple linear layer module for acoustic signal classification. It achieved a training accuracy of 96.12\% and a testing accuracy of 93.98\%, which verifies the newly generated dataset’s quality. Shown in Table \ref{tab3} is the summary on the comparison of the original dataset with the new dataset.

\begin{table}[tbp]
\caption{Combining and feature extraction methods}
\begin{center}
\begin{tabular}{|p{0.27\columnwidth}|p{0.28\columnwidth}|p{0.27\columnwidth}|}
\hline
\textbf{Combining \newline method} & \textbf{Feature \newline extraction \newline method} & \textbf{Result} \\
\hline
Extending & Time-seires & Failed \\
\hline
Extending & Spectrogram & Failed \\
\hline
Extending & MFE & Failed \\
\hline
Extending & MFCC & Failed \\
\hline
Summing & Time-seires & Failed \\
\hline
Summing & Spectrogram & Failed \\
\hline
Summing & MFE & Failed \\
\hline
Summing & MFCC & Successful \\
\hline
\end{tabular}
\label{tab1}
\end{center}
\end{table}

\begin{table}[tbp]
\caption{Deformation and filters used for data augmentation}
\begin{center}
\begin{tabular}{|p{0.43\columnwidth}|p{0.43\columnwidth}|}
\hline
\textbf{Deformation techniques \newline \textit{(Audiomentation)}} & \textbf{Frequency filters\newline \textit{(Procedural generation)}} \\
\hline
stretch\_factor = \newline random.uniform(0.8,1.2) \newline shift\_factor = random.\newline randint(-1000,1000) \newline pitch\_factor = random.\newline randint(-3,3) \newline compression\_factor = \newline random.uniform(0.1,0.5) \newline noise\_factor = random. \newline uniform(0,0.05) \newline shift\_factor = random. \newline uniform(-0.1,0.1) \newline filter\_factor = random. \newline randint(10,90) \newline audio\_sep = harmonic\_ \newline percussive\_separation \newline ((audio,sr)) \newline audio\_vibrato = \newline vibrato((audio,sr)) & applying\_time\_varying\_ \newline lowpass\_filter \newline (audio\_data, sr): \newline filter\_order = 6 \newline window\_size = 1024 \newline overlap = 0.5 \newline butter\_lowpass \newline (cutoff, fs, order=5): \newline nyq = 0.5 * fs \newline normal\_cutoff = cutoff / nyq \newline butter\_lowpass\_filter \newline (data, cutoff, fs, order=5): \newline b,a = butter\_lowpass \newline (cutoff, fs, order=order) \\
\hline
\end{tabular}
\label{tab2}
\end{center}
\end{table}

\begin{table}[tbp]
\caption{Samples per maturity level (original vs newly generated dataset)}
\begin{center}
\begin{tabular}{|p{0.24\columnwidth}|p{0.25\columnwidth}|p{0.30\columnwidth}|}
\hline
\textbf{Maturity \newline level} & \textbf{Original \newline dataset \cite{15}} & \textbf{Newly generated \newline dataset} \\
\hline
Premature & 24 & 4,050 \\
\hline
Mature & 108 & 4,050 \\
\hline
Overmature & 255 & 5,850 \\
\hline
\textbf{Total} & \textbf{387} & \textbf{13,950} \\
\hline
\end{tabular}
\label{tab3}
\end{center}
\end{table}

\begin{figure}[tbp]
\centering
\includegraphics[scale=0.26]{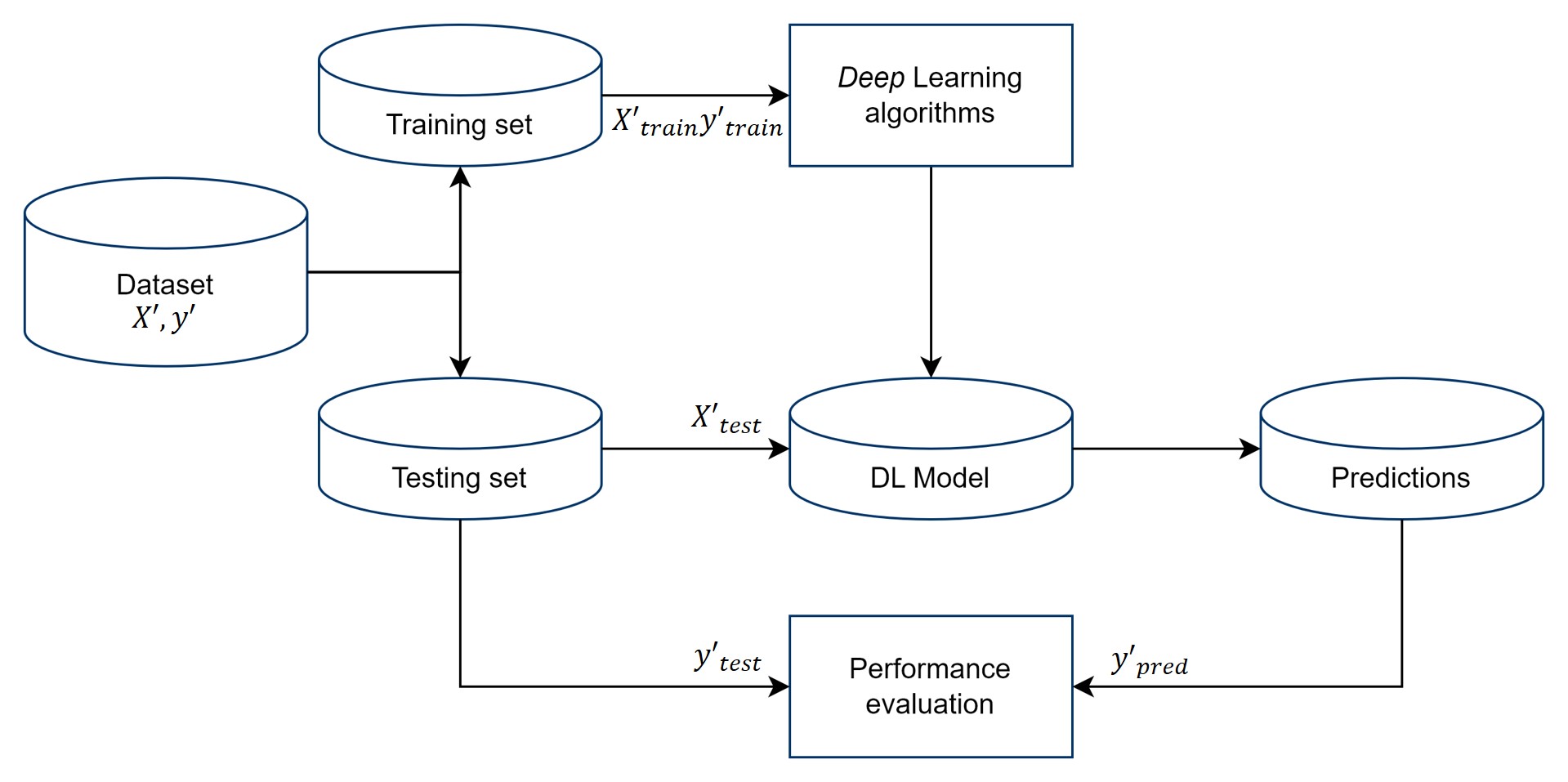}
\caption{A deep learning pipeline with training and testing phases}
\label{fig_1}
\end{figure}

\subsection{Deep learning model}

DL is a method of analyzing data for automating inference based on historical data; a deep hierarchical extension of machine learning \cite{24}. In Fig. \ref{fig_1}, a DL pipeline with training and testing phases and performance evaluation is shown. Before the dataset was fed to the learning algorithms, the dataset has to be divided into training and testing sets \cite{25}. Similar to machine learning, DL is also composed of two major phases—training and testing phases. Training and testing phases are fundamental steps in designing a model, especially in machine and deep learning \cite{26}. The difference between the two is that DL can ingest unstructured data in its raw form while machine learning cannot \cite{27}. This makes DL superior to machine learning when it comes to processing datasets.

\begin{figure*}[tbp]
\centering
\includegraphics[width=15cm, height=8cm]{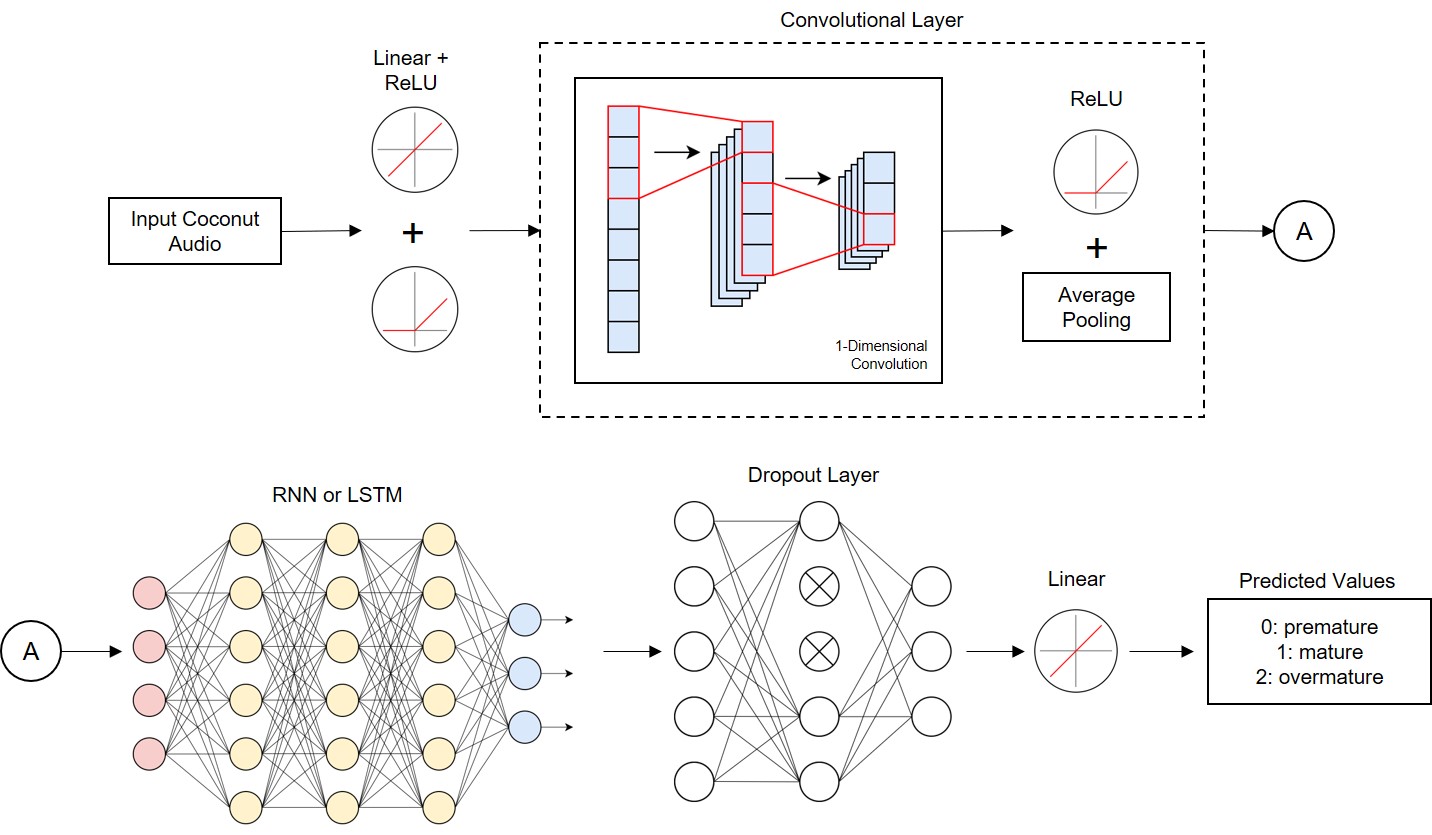}
\caption{Proposed model architecture}
\label{fig_2}
\end{figure*}

\begin{table}[tbp]
\caption{Parameters used in the model architecture}
\begin{center}
\begin{tabular}{|p{0.35\columnwidth}|p{0.50\columnwidth}|}
\hline
\textbf{Layer and model} & \textbf{Parameters}\\
\hline
Convolutional layer & nn.Conv1d(128, 32, 3, \newline padding=1) \newline nn.Conv1d(32, 64, 3, \newline padding=1) \newline nn.AvgPool1d(1) \\
\hline
LSTM model & hidden units each layer = 64 \newline batch size = 128 \newline epoch = 60 \newline learning rate = 0.001 \\
\hline
RNN model & hidden units each layer = 64 \newline batch size = 128 \newline epoch = 60 \newline learning rate = 0.001 \\
\hline
Dropout layer & nn.Dropout(0.5)\\
\hline
\end{tabular}
\label{tab4}
\end{center}
\end{table}

\subsection{Model architecture development}

The dataset with the augmented data is used as data inputs to two prominent time series DL models, which are RNN and LSTM. The proposed model architecture is shown in Fig. \ref{fig_2}, in which the dataset will undergo the same layers but with different parameters and DL models in their respective training and testing phases. The architecture is composed of the following: linear and rectified linear unit (ReLU) activation functions, a convolutional layer, a DL model of either RNN or LSTM, a dropout layer, and an output layer. All parameters used in the layers of the architecture are summarized in Table \ref{tab4}.

The convolutional layer with a pooling layer, an average pooling layer in the proposed model architecture, is connected to the architecture to perform feature extraction on the input signal \cite{28}. The one-dimensional convolutional layer can effectively extract features through convolution calculation \cite{29}. The acoustic signals taken from coconut knocks are considered one-dimensional data, thus, the one-dimensional convolutional layer is applicable to this study. The feature to be extracted in this layer is Mel Frequency Cepstral Coefficient (MFCC). MFCC is characterized as a compact representation of the spectrum of an audio signal that contains information about rate changes in different spectrum bands \cite{30}. The convolutional layer is implemented twice with different parameters in this architecture then data inputs are processed in the fully-connected layer.

RNN is a DL neural network that specializes in dealing with sequential and time series data, which is commonly used for ordinal or temporal problems \cite{31}. Its architecture possesses cycles and allows previous outputs to be used as inputs while having hidden states \cite{32}. RNN does not assume that inputs and outputs are independent but rather, the output of the model depends on its prior element within the sequence \cite{31}. On the other hand, the LSTM network is a special class of network under the class of RNN, in which it also deals with sequential and time series data and its connection contains at least one cycle \cite{20}. It differs from RNN in such a way that the memory cells in its hidden layers are composed of three gates: forget, input, and output gates \cite{20}. These gates allow an LSTM memory cell to access and store information over long periods of time \cite{33}. Both models have been successfully applied in numerous studies dealing with such a nature of datasets \cite{34}.

The dropout layer adopts a widely used technique, called dropout, in which features created by prior layers are redundant \cite{35}. In other words, specific nodes are being deactivated by setting them to zero during the training phase \cite{36}. This is to avoid co-adaptation of feature detectors or overfitting \cite{37}. The last layer is the output layer in which the model will predict the maturity levels of the signals. The three maturity levels are coded to their respective numerical values (0: “premature,” 1: “mature,” 2: “overmature”).

\subsection{Classification performance evaluation}

Commonly, datasets are divided into training and testing sets \cite{38}. The dataset partition is set at 90/10, wherein 90\% of the dataset was used for the training phase while the remaining 10\% is for the testing phase. Furthermore, a \(k\)-fold cross-validation was applied, wherein both training and testing sets were taken from the same dataset [24]. The \(k\)-fold cross-validation is the most widely adopted estimator among many estimators of the prediction error of a classifier model \cite{39}. In practice, \(k\) between 5 and 10 is most typically used, but Burman \cite{40} reasoned that \(k>5\) might cause an issue.
The two DL models, RNN and LSTM, will be compared based on performance indicators. Following Agarwal et al. \cite{41}, these are the metrics used: accuracy, precision, recall, and F1-score. Accuracy specifies the fraction of times that the model predicts correctly. Precision is the proportion of all predictions that were made by the model that is actually true. Recall is the fraction of times that the classification model can identify all relevant instances. Lastly, F1-score is a single metric that represents the harmonic mean between recall and precision values. All metrics are calculated as

\[Accuracy=\frac{TP+TN}{TP+FP+FN+TN}\]
\[Precision=\frac{TP}{TP+FP}\]
\[Recall=\frac{TP}{TP+TN}\]
\[F1-score=2\left( \frac{1}{\frac{1}{Precision}+\frac{1}{Recall}} \right)\]

where TP is true positive, TN is true negative, FP is false positive, and FN is false negative.

\section{Results and Discussions}

\begin{figure*}[tbp]
\centering
\includegraphics[width=15cm, height=7cm]{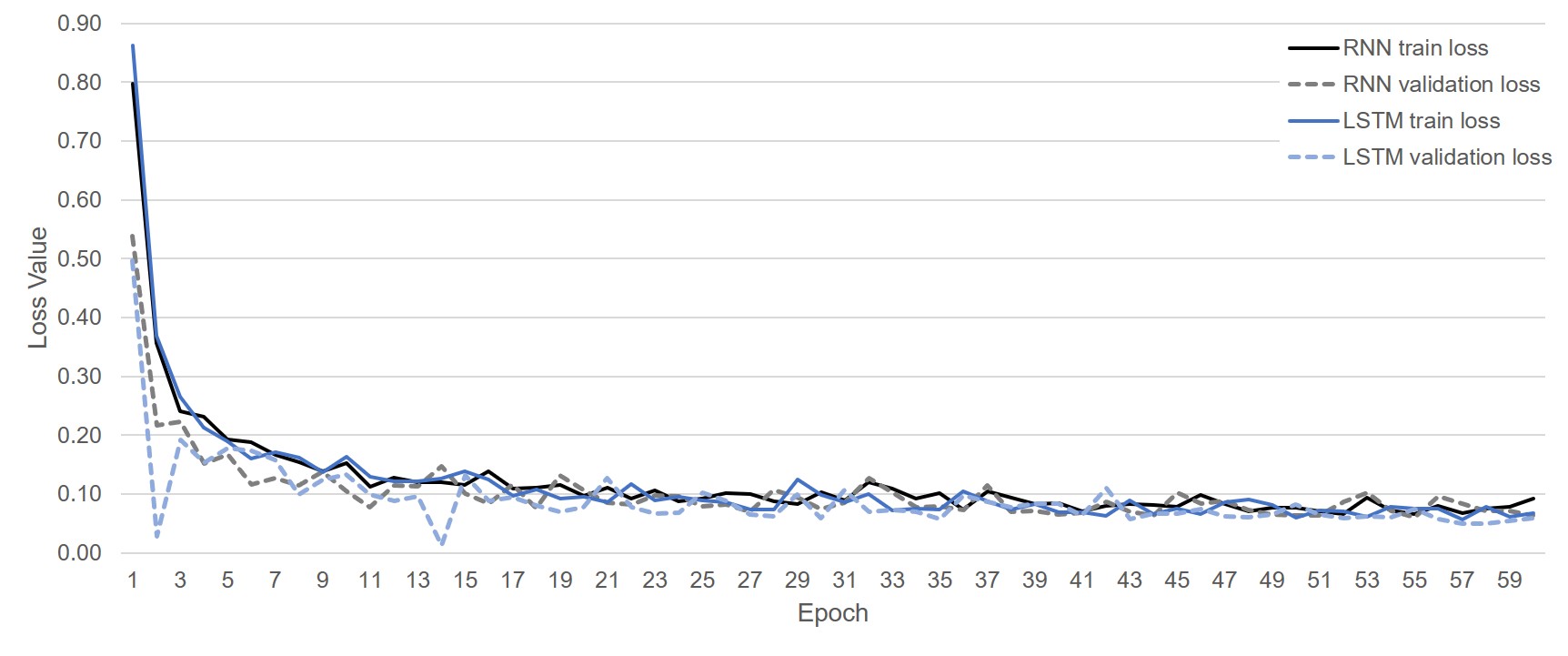}
\caption{Loss graph for RNN and LSTM models}
\label{fig_3}
\end{figure*}

\begin{figure}[tbp]
\centering
\includegraphics[scale=0.17]{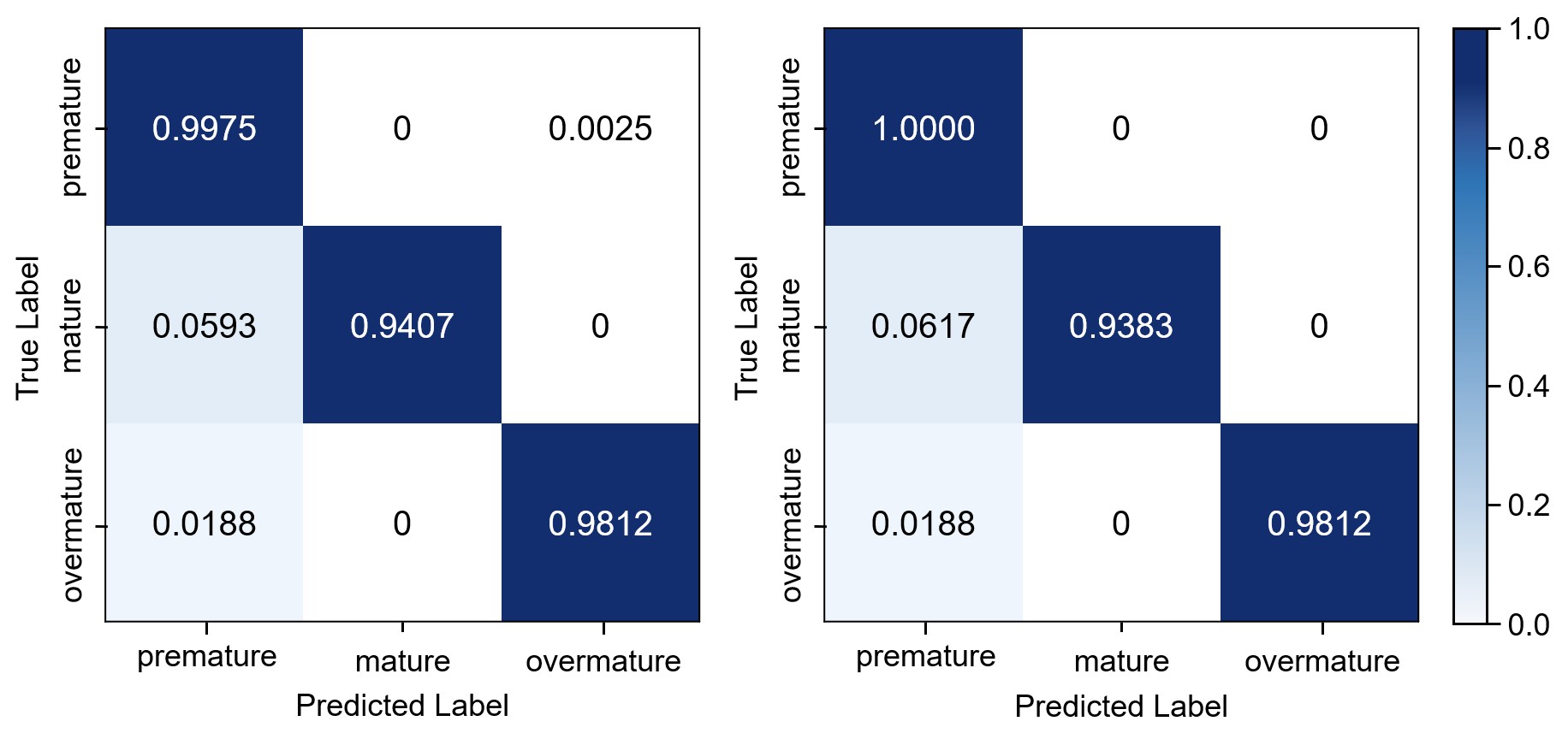}
\caption{Confusion matrix for RNN model(left) and LSTM model (right)}
\label{fig_4}
\end{figure}

After generating augmented data to be added to the dataset and establishing the model architecture, RNN, and LSTM models are trained, tested, and evaluated in classifying acoustic signals to their maturity levels. The loss function graph for the two models is shown in Fig. \ref{fig_3}. Both loss values of the models did not converge to zero but are closely approaching zero. There are also no signs of overfitting since the validation loss values remain less than train loss values under both models. RNN achieved its lowest loss value at epoch 52 with 0.0657 while LSTM has its lowest loss value at epoch 57 with 0.0572.

Presented in Fig. \ref{fig_4} is the confusion matrix of the two models. For the premature level, LSTM correctly predicted all samples under this maturity level while RNN correctly predicted 99.75\%, which is 0.25\% lesser than LSTM. Moreover, RNN correctly predicted 94.07\% under the mature level compared to LSTM having 93.83\%; 0.24\% less than RNN. Lastly, both models correctly predicted 98.12\% under the overmature level. Both models are classified less under the mature level compared to the other two maturity levels, which are almost, if not, 100\%. It can be observed that RNN and LSTM perform at equal levels looking at the percentages of correct predictions.

To further elaborate on the classification performance of the two models, presented in Table \ref{tab4} is a comparison of the two models using the performance indicators discussed in Section 2.5. Both models have equal accuracy and average recall. As for the average precision and F1-score, RNN is greater than LSTM by 0.05\% and 0.02\%, respectively. Despite these differences, RNN and LSTM still perform almost similarly. When checking if the percentages from all performance evaluations have any significant difference, the p-value arrived at 0.9930, which is greater than 0.05, thus, there is no significant difference between the results of the two models.

One of the objectives of the study is to improve the performance accuracy of the models from the study of Caladcad et al. \cite{17}. To evaluate if the results from the two DL models have improved, their classification performance is compared to the results of the ML models used in the study of Caladcad et al. \cite{17}. A summary of this is presented in Table \ref{tab5}. There is an obvious difference in the results of the percentages on correct predictions between DL and ML models. Only in classifying overmature samples are the ML models greater than DL models but the difference is only at 1.88\%, which is not much. The difference in the results for the premature level ranges from 61.75\% to 75\% and in the mature level, it ranges from 34.83\% to 56.07\%. For accuracy, the difference between DL and ML models ranges from 13.94\% to 17.42\%. On the other hand, the difference in the F1-scores ranges from 15.85\% to 20.55\%.

Checking if the performance evaluation results between ML and DL models have any significant differences, the p-value is 0.001, which is less than 0.05, thus, there is a significant difference between the results of the two types of models. DL models clearly outperformed the ML models. The biggest contributor to this is the amount of the dataset processed by the DL models while the results from ML models were taken from an imbalanced dataset. Since this was resolved by data augmentation before the dataset was processed by the DL models, it significantly improved the classification performance. With this, the performance of the models relies on the dataset they processed.

The study further proves the application of DL models in the field of agriculture. The results of this study can be integrated into a designed system for coconut maturity classification in proposal for a non-destructive classification system for coconut mass exportation. This can help agricultural industry lessen cost related to fruit loss and classification inefficiencies and inaccuracies, and increase productivity. Although hardware development will be a challenge, most especially for large-scale operations, since it’ll be costly and more simulations and tests shall be conducted prior to actually implementing such technology in real-world.

\begin{table}[tbp]
\caption{Performance comparison of the two DL models}
\begin{center}
\begin{tabular}{|p{0.27\columnwidth}|p{0.28\columnwidth}|p{0.27\columnwidth}|}
\hline
\textbf{Performance \newline indicators} & \textbf{Recurrent \newline neural network} & \textbf{Long short- \newline term memory} \\
\hline
Accuracy (\%) & 97.42 & 97.42 \\
\hline
Average \newline precision (\%) & 97.28 & 97.23 \\
\hline
Average recall \newline (\%) & 97.32 & 97.32 \\
\hline
F1-score (\%) & 97.22 & 97.20 \\
\hline
\end{tabular}
\label{tab5}
\end{center}
\end{table}

\begin{table}[tbp]
\caption{Performance comparison of DL models and ML models from the study of Caladcad et al. \cite{15}}
\begin{center}
\begin{tabular}{|p{0.27\columnwidth}|p{0.27\columnwidth}|p{0.27\columnwidth}|}
\hline
\textbf{Performance \newline evaluation} & \textbf{Accuracy (\%)} & \textbf{F1-score (\%)} \\
\hline
RNN & 97.42 & 97.22 \\
\hline
LSTM & 97.42 & 97.20 \\
\hline
ANN \cite{15} & 81.74 & 79.27 \\
\hline
RF \cite{15} & 83.48 & 81.35 \\
\hline
SVM \cite{15} & 80.00 & 76.67 \\
\hline
\end{tabular}
\label{tab6}
\end{center}
\end{table}

\section{Conclusion and future works}

This study utilizes the advancement of technology and the rise of smart applications to propose a mechanical method to the traditional methods of classifying coconut fruit for exportation, which are prone to inconsistencies due to human error. It also extends the study of Caladcad et al. \cite{17} by addressing the following gaps: (1) the imbalanced dataset, (2) the need to update the classification system, and (3) the performance accuracy of the MLAs used in the study. RNN and LSTM models have classification performances similar to each other and it can be concluded that neither RNN nor LSTM outperformed each other, making either of the models suitable for classifying coconuts to their maturity levels. However, when the DL and ML models were compared, the DL models outperformed the ML models. The increase in the data inputs in the dataset processed by the DL models greatly contributed to the huge improvement of the classification performance. Thus, it can also be concluded that the model performance heavily relies on the quality and amount of dataset it has processed. For future works, the study can still be improved by addressing the limited dataset prior to data augmentation. More samples can be gathered to enhance preprocessing and consequently, improve models’ performance. Other than acoustic signals, other features such as physicochemical properties of coconut can be explored to even enhance classification performance.

\renewcommand\refname{References}
\bibliographystyle{unsrt} 
\bibliography{coconut_DL_IEEEHTC.bib}
\vspace{12pt}

\end{document}